\documentclass[apjl]{myemulateapj}

\usepackage{comment}
\usepackage{graphicx}
\usepackage{ifpdf}
\usepackage{ifthen}

\newboolean{emulateapj}
\setboolean{emulateapj}{true}

\newcommand{\ctbd}[1]{}

\newcommand{\lc}{light curve}
\newcommand{\lcs}{light curves}

\newcommand{\cfa}{Harvard-Smithsonian Center for Astrophysics (CfA)}

\newcommand{\pxs}{\ensuremath{\rm \arcsec pixel^{-1}}}

\newcommand{\tsize}[1]{\mbox{\rm #1 m}}

\newcommand{\kms}{\ensuremath{\rm km\,s^{-1}}}
\newcommand{\ms}{\ensuremath{\rm m\,s^{-1}}}

\newcommand{\teff}{\ensuremath{T_{\rm eff}}}
\newcommand{\logg}{\ensuremath{\log{g}}}
\newcommand{\vsini}{\ensuremath{v \sin{i}}}
\newcommand{\feh}{[Fe/H]}
\newcommand{\logl}{\ensuremath{\log{L}}}

\newcommand{\rsun}{\ensuremath{R_\sun}}
\newcommand{\msun}{\ensuremath{M_\sun}}
\newcommand{\lsun}{\ensuremath{L_\sun}}

\newcommand{\mstar}{\ensuremath{M_*}}

\newcommand{\loglstar}{\ensuremath{\log{L_*}}}

\newcommand{\rpl}{\ensuremath{R_{P}}}
\newcommand{\mpl}{\ensuremath{M_{P}}}

\newcommand{\ipl}{\ensuremath{i_{P}}}

\newcommand{\rjup}{\ensuremath{R_{\rm J}}}
\newcommand{\mjup}{\ensuremath{M_{\rm J}}}

\newcommand{\msini}{\ensuremath{m \sin i}}

\newcommand{\pack}[1]{\textsc{\lowercase{#1}}}
\newcommand{\prog}[1]{\texttt{\lowercase{#1}}}

\newcommand{\fihat}{\pack{fihat}}

\newcommand{\fiphot}{\prog{fiphot}}

\newcommand{\figr}[1]{Fig.~\ref{fig:#1}}
\newcommand{\secr}[1]{\mbox{\S\ \ref{sec:#1}}}

\newcommand{\tabr}[1]{\mbox{Table~\ref{tab:#1}}}

\newcommand{\ads}{\mbox{ADS 16402}}
\newcommand{\adsa}{\mbox{ADS 16402A}}
\newcommand{\adsb}{\mbox{ADS 16402B}}
\newcommand{\adsab}{\mbox{ADS 16402AB}}
\newcommand{\hatp}{HAT-P-1b}

\shorttitle{Large Radius Transiting Planet in a Stellar Binary}
\shortauthors{Bakos et al.}

\begin{document}
\ifthenelse{\boolean{emulateapj}}{
\title{HAT-P-1\lowercase{b}: A Large-Radius, Low-Density Exoplanet Transiting
	one Member of a Stellar Binary\altaffilmark{$\dagger$ $\star$}}}
{\title{HAT-P-1b: A Large-Radius, Low-Density Exoplanet Transiting
	one Member of a Stellar Binary\altaffilmark{\dagger\, \star}}}

\author{
	G.~\'A.~Bakos\altaffilmark{1,2},
	R.~W.~Noyes\altaffilmark{1},
	G.~Kov\'acs\altaffilmark{3},
	D.~W.~Latham\altaffilmark{1},
	D.~D.~Sasselov\altaffilmark{1},
	G.~Torres\altaffilmark{1},
	D.~A.~Fischer\altaffilmark{6},
	R.~P.~Stefanik\altaffilmark{1},
	B.~Sato\altaffilmark{7},
	J.~A.~Johnson\altaffilmark{8},
	A.~P\'al\altaffilmark{4,1},
	G.~W.~Marcy\altaffilmark{8},
	R.~P.~Butler\altaffilmark{9},
	G.~A.~Esquerdo\altaffilmark{1},
	K.~Z.~Stanek\altaffilmark{10},
	J.~L\'az\'ar\altaffilmark{5},
	I.~Papp\altaffilmark{5},
	P.~S\'ari\altaffilmark{5} \&
	B.~Sip\H{o}cz\altaffilmark{4,1}
}

\altaffiltext{1}{\cfa,
	60 Garden Street, Cambridge, MA 02138, USA; gbakos@cfa.harvard.edu.}
\altaffiltext{2}{Hubble Fellow.}
\altaffiltext{3}{Konkoly Observatory, Budapest, P.O.~Box 67, H-1125, Hungary}
\altaffiltext{4}{Department of Astronomy,
	E\"otv\"os Lor\'and University, Pf.~32, H-1518 Budapest, Hungary.}
\altaffiltext{5}{Hungarian Astronomical Association, 1461 Budapest, 
	P.~O.~Box 219}
\altaffiltext{6}{Department of Physics \& Astronomy, San Francisco
	State University, San Francisco, CA 94132, USA;
	fischer@stars.sfsu.edu}
\altaffiltext{7}{Okayama Astrophysical Observatory, National
	Astronomical Observatory, Kamogata, Asakuchi, Okayama 719-0232,
	Japan}
\altaffiltext{8}{Department of Astronomy, University of California,
	Berkeley, CA 94720, USA}
\altaffiltext{9}{Department of Terrestrial Magnetism, Carnegie  
	Institute of Washington DC, 5241 Broad Branch Rd.~NW, Washington
	DC, USA 20015-1305}
\altaffiltext{10}{Department of Astronomy, Ohio State University, 140 West
	18th Avenue, Columbus, OH 43210}

\altaffiltext{$\dagger$}{
	Based in part on data collected at the Subaru Telescope, which is
	operated by the National Astronomical Observatory of Japan.
}
\altaffiltext{$\star$}{
	Based in part on observations obtained at the W.~M.~Keck
	Observatory, which is operated by the University of California and
	the California Institute of Technology. Keck time has been granted
	by NASA.
}
\setcounter{footnote}{1}

\begin{abstract}

Using small automated telescopes in Arizona and Hawaii, the HATNet
project has detected an object transiting one member of the double star
system \ads. This system is a pair of G0 main-sequence stars with age
about 3 Gyr at a distance of $\sim$139 pc and projected separation of
$\sim$1550 AU.  The transit signal has a period of 4.46529 days and
depth of 0.015 mag. From follow-up photometry and spectroscopy, we find
that the object is a ``hot Jupiter'' planet with mass about 0.53 \mjup\
and radius $\sim$1.36 \rjup\
traveling in an orbit with semimajor axis 0.055 AU and
inclination about 85\fdg9,
thus transiting the star at 
impact parameter 0.74 of the stellar radius. Based on a data set
spanning three years, ephemerides for the transit center are: 
$T_C = 2453984.397 + N_{tr} * 4.46529$.
The planet, designated \hatp, appears to be at least as large in
radius, and smaller in mean density, than any previously-known planet.
\end{abstract}

\keywords{
	stars: individual: {\mbox ADS 16402 AB} \---
	planetary systems: individual: {HAT-P-1b} \---
	binaries}

\section{Introduction and Summary}
\label{sec:intro}
The 10 currently known transiting extrasolar planets occupy a special
place among the nearly 200 known exoplanets because their photometric
transits yield unambiguous information on their masses and radii.
Five of these transiting planets were found through spectroscopic and
photometric follow-up of candidates discovered by the OGLE project
\citep[these are OGLE-TR-10b, 56b, 111b, 113b, and 132b; ]
[respectively]{udalski02a,udalski02b,udalski02c,udalski03,udalski04}. 
For references to the follow-up confirmation papers for these objects,
see the Extrasolar Planets 
Encyclopedia\footnote{http://exoplanet.eu/}.
The parent stars of these planets are all rather faint, making
effective followup somewhat difficult.
The other 
five\footnote{Discovery of {\mbox TrES-2} \citep{odonovan06} 
was announced during the refereeing of this paper.}
transiting exoplanets orbit nearby, bright stars:
 {\mbox HD 209458b}
 \--- \citet{dc00,henry00},
 {\mbox TrES-1}
 \--- \citet{alonso04},
{\mbox HD 149026b} 
 \--- \citet{sato05},
 {\mbox HD 189733b}
 \--- \citet{bouchy05},
 {\mbox XO-1b} 
 \--- \citet{mccullough06}.
These planets are of special interest, as accurate parameter
determination as well as other types of follow-up are facilitated.  For
this reason, a number of wide-field surveys, using small telescopes,
are underway \citep[see, e.g. ][]{dc06b}.
Indeed, two of the 5 known transiting planets around nearby stars, 
{\mbox TrES-1} and 
{\mbox XO-1b}, were detected by such surveys.
(The other three were first detected by radial velocity measurements).

The HATNet project,\footnote{www.hatnet.hu}
initiated in 2003 by G.~\'A.~B, is also a
wide-field survey that aims for the discovery of transiting planets around
bright stars. It currently comprises 6 small wide-field automated
telescopes, each of which monitors $8\arcdeg\times8\arcdeg$ of sky,
typically containing 5000 stars bright enough to permit detection of
planetary transits through the typically 1\% photometric dips they
induce on their parent stars. The instruments are spread in a
two-station, longitude-distributed network, with four telescopes at the
F.~L.~Whipple Observatory (FLWO) in Arizona, and two telescopes at the
Submillimeter Array (SMA) at Hawaii. Technical aspects of HATNet will
be described in a forthcoming paper, but for instrumentation,
observations, and data flow, see \citet{bakos02,bakos04}. Here we
report on the first detection by HATNet of a transiting extrasolar
planet. This planet, which we hereafter label as \hatp, orbits a 10th
magnitude star, and thus is well-suited for follow-up with large
ground-based and space telescopes.

\notetoeditor{This is the intended place of \figr{stamps}}

\ifthenelse{\boolean{emulateapj}}{\begin{figure*}[t]}{\begin{figure}[t]}
\ifpdf
	\plottwo{f1a.pdf}{f1b.pdf}
\else
	\plottwo{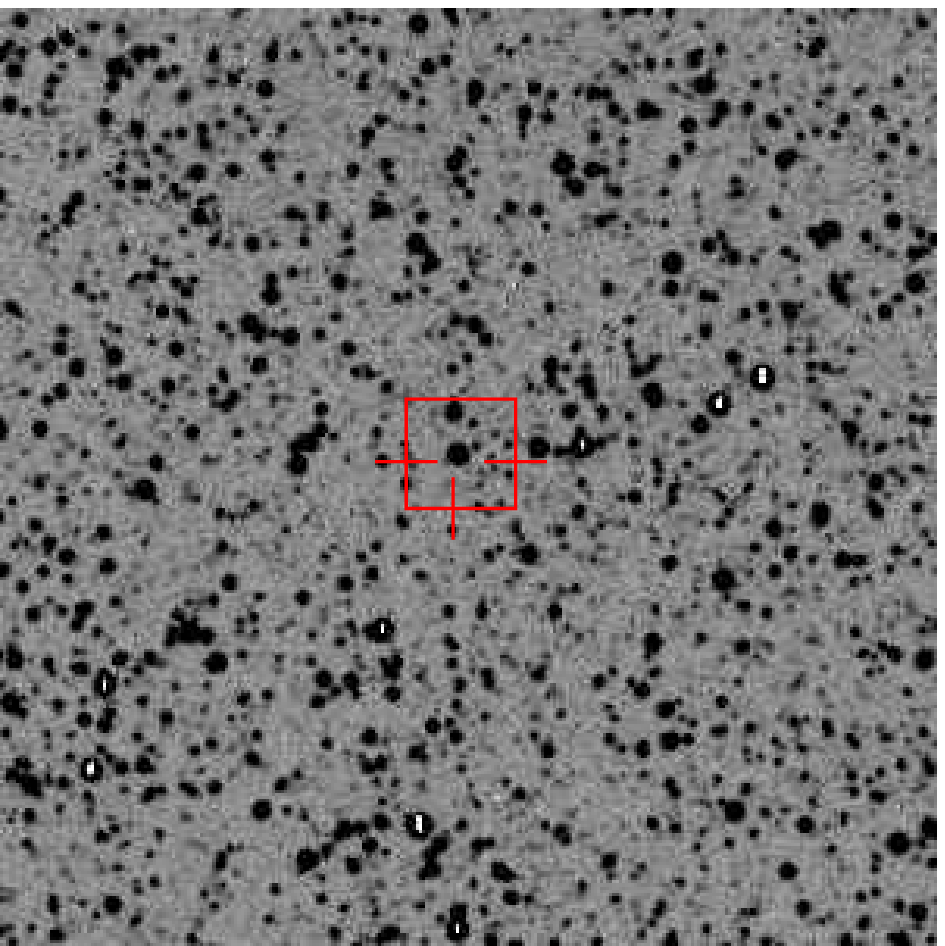}{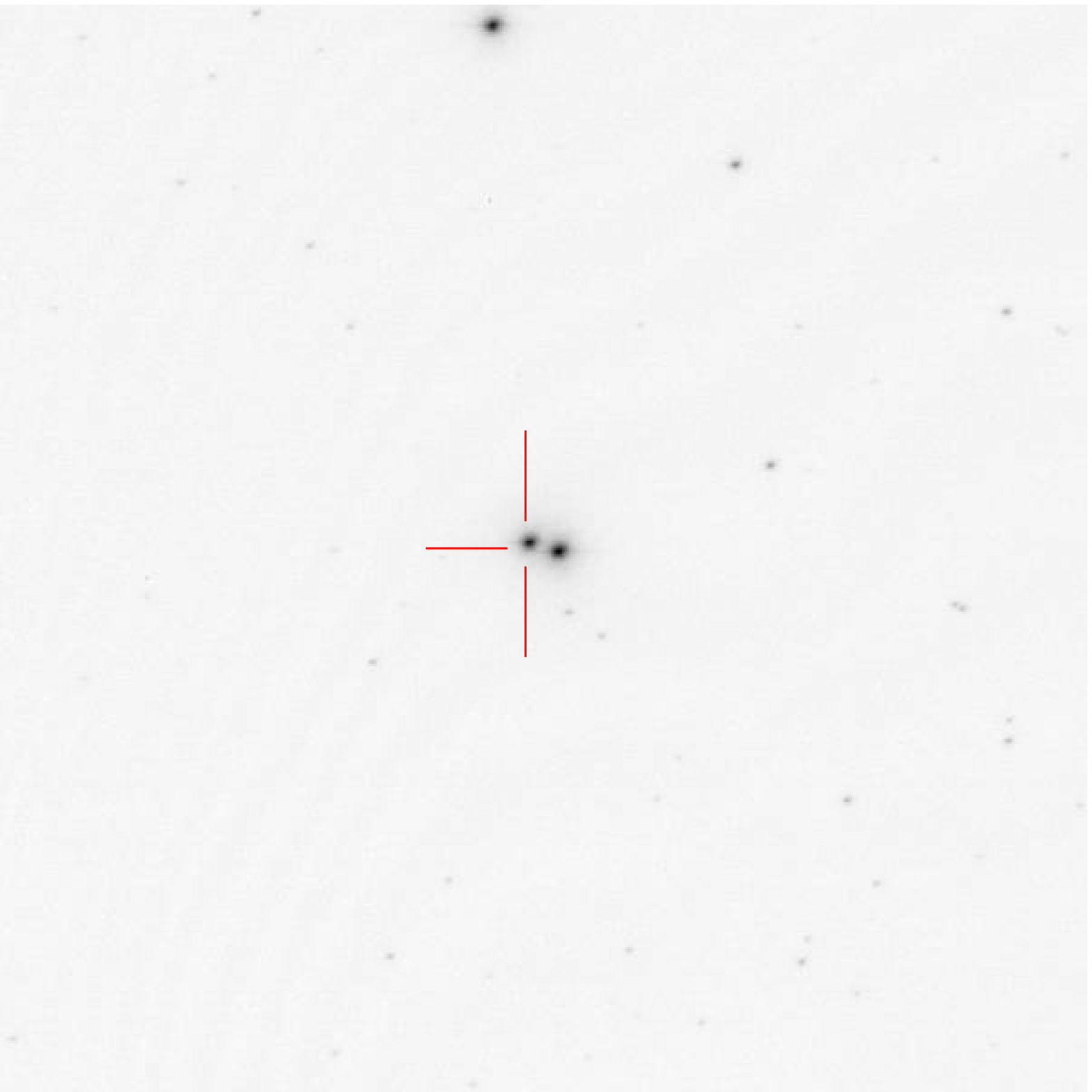}
\fi
\caption{
	The left panel shows a $70\arcmin\times70\arcmin$ (300 pixels wide) stamp
	of a typical image taken by HAT-5. The stamp width is about 1/7th of the
	original frame. \ads\ is in the middle (the two components are merged
	into a bright star). Peaks of saturated stars are masked, and shown in
	white.  The right panel is a stamp from an image taken with the FLWO
	\tsize{1.2} telescope. It is $6\farcm8\times6\farcm8$ (600 pixels wide), and
	corresponds to the small box on the left panel. \adsb (marked with the
	cross) is well separated from \adsa.
\label{fig:stamps}}
\ifthenelse{\boolean{emulateapj}}
{\end{figure*}}{\end{figure}}

The existence of \hatp\ was first inferred from a 0.6\% (6 mmag) deep
transit-like dip seen in the $I$-band \lc\ of the stellar system \ads\
\citep[Aitken Double Star Catalogue; ][]{aitken32} obtained by the
combination of the HAT-5 instrument at FLWO and HAT-8 at the SMA. The
members of this binary stellar system are a pair of nearly identical
G0V stars, \adsa\ ($I\approx9$) and \adsb\ ($I\approx9.6$) separated
by 11\farcs2.  The system appears as a single, elongated source in
the HAT images because this separation is less than the 14\pxs\ image
resolution.  Initial follow-up spectroscopy at the FLWO \tsize{1.5} telescope
showed the absence of radial velocity variations of either star at a
level of 1 \kms\ or larger, demonstrating that the source of the
photometric dips could not be due to a common false positive, namely an
M dwarf star transiting in front of one of the stars. Subsequent
spectroscopy at the Subaru telescope using N2K project observing time
\citep{fischer04,sato05}, and later at the W.~M.~Keck telescope
revealed sinusoidal velocity variations of \adsb\ with an amplitude of
60\ms, thus suggesting the existence of a 0.5\mjup\ planet transiting
that star. After excluding blend scenarios, the combination of the
photometric and spectroscopic data leads to the conclusion that this
object, \hatp, is indeed a planet, and has a radius comparable to that
of {\mbox HD 209458b}, and a somewhat lower mean density. In the
remainder of this paper we discuss all of the above points in more
detail.

\section{Observations}
\label{sec:Obs}

\subsection{HATNet Discovery Photometry} 

\ads\ (also {\mbox BD+37 4734},
$\alpha = 22^h57^m46\fs8$, 
$\delta = +38\arcdeg40\arcmin28\arcsec$, J$2000.0$)
is a visual double star system that also bears the name HJ1832, as it
was discovered by John Herschel
\citep[][]{herschel31}. The system
lies in HATNet survey field G205 
(centered at $\alpha = 22^h55^m$, 
$\delta = +37\arcdeg30\arcmin$).
2059 observations were made of this field by the HAT-5 telescope and
901 by HAT-8 between 2003 October 13 and 2004 January 30. A small stamp
taken from a typical HATNet image is shown on the left panel of
\figr{stamps}. Exposure times were 5 min at a cadence of 5.5 min. Light
curves were derived by aperture photometry for the 6400 stars in G205
bright enough to yield photometric precision of better than 2\%
(reaching 0.3\% in some cases).  In deriving the \lcs, we made use of
the Trend Filtering Algorithm \citep[TFA;][]{kovacs05} to correct for
spurious trends in the data.
This was particularly important for the \lc\ for \ads, for which
variable blending distorts the combined shape of the two close-lying
unresolved components; indeed the shallow transit signal in the
combined light of the two stars is only barely visible in the raw \lc,
while it is readily seen after application of TFA.

We then searched all the \lcs\ for characteristic transit signals,
using the Box-fitting Least Squares \citep[BLS; ][]{kovacs02}
algorithm, which seeks for box-shaped dips in the parameter space of
frequency, transit duration, and phase of ingress. Candidate transit
signals with the highest detection significance were then examined
individually, to isolate those with the best combination of stellar
type (preferably main-sequence stars of spectral type mid-F or later),
and depth, shape, and duration of transit.  One of these was \ads, for
which we identified a prominent boxcar-like periodicity with period
4.4656 days and transit depth in the combined light of the two stars of
6mmag.
Portions of five transits were observed during the observing interval.
The upper panel of \figr{hat_flwo48_lc} shows the combined data for
these five transits, with different symbols for observations from HAT-5
(Arizona) and HAT-8 (Hawaii).

\subsection{Rejection-mode Spectroscopic Observations}
\label{sec:rejspec}

Initial follow-up observations were made with the CfA Digital
Speedometer \citep[DS;][]{latham92} in order to characterize the two
individual stars comprising \ads, as well as to search for evidence
that the transit signal could be due to the transit of a late M dwarf
in front of either of the stars, as has been found for many otherwise
plausible planet transit candidates in the HATNet database. These
observations indicated that both sources indeed have values of
\teff\ and \logg\ consistent with main-sequence late F or early G
dwarfs. The DS data revealed no sign of radial velocity variation in
either of the two components, with upper limit of detection about
1.7\kms. For a period of $\sim$4.5d this limit corresponds to a
secondary mass of 15\mjup, thus ruling out the possibility that the
transit signal is due to a low-mass stellar companion transiting one of
the two stars.

\subsection{Follow-up Photometry Observations}
\label{sec:photfol}

\begin{figure}[h]
\ifpdf
	\plotone{f2.pdf}
\else
	\plotone{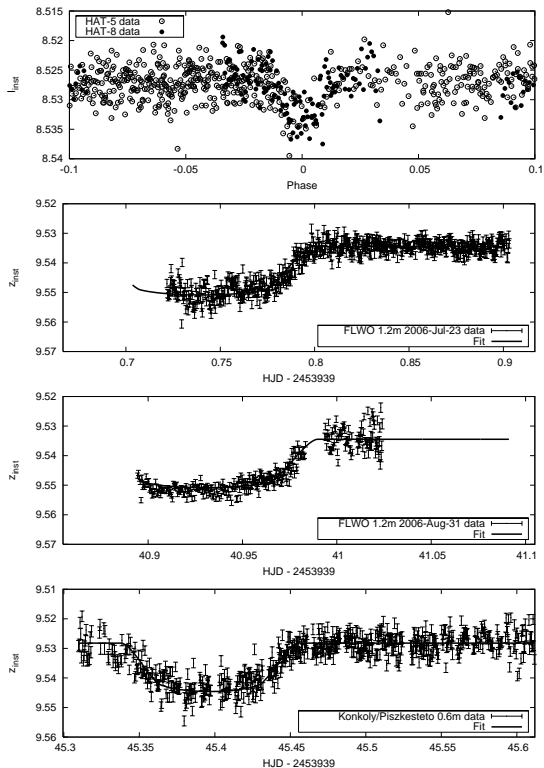}
\fi
\caption{
	The upper panel shows the HAT-5 (open circles) and HAT-8 (filled
	circles) TFA corrected \lc\ from FLWO, Arizona and SMA, Hawaii,
	respectively.  Altogether five transit events were observed
	successfully between the two sites. The data are phased at period
	P=4.46529 days.  Zero phase corresponds to the time of mid-transit
	using the ephemeris discussed in the text. The second and third
	panels show the follow-up observations taken with KeplerCam on the
	FLWO \tsize{1.2} telescope in $z$-band in 2006 July 23 and 2006 Aug 31,
	respectively.  Unfortunately only the central and egress portions
	of the transit were observable.  The bottom panel shows the
	$I$-band data of a full-transit taken by the Konkoly Observatory
	0.6 m Schmidt telescope at 2006 Sep 5. Superimposed on the lower
	three panels are our best fits (\secr{PlanetProperties}) to the
	transit \lc.
\label{fig:hat_flwo48_lc}}
\end{figure}

\notetoeditor{This is the intended place of \figr{hat_flwo48_lc}}

After identification of the \ads\ system as a candidate for harboring a
transiting extrasolar planet, we used the TopHAT telescope to search
for evidence of a transit. TopHAT is a \tsize{0.26} aperture automated
telescope at FLWO, part of the HATNet, designed for follow-up of
transit candidates \citep{bakos06}. With its pixel scale of 2.2\pxs\,
it is able to resolve the two stars in the system. However, data were
taken during the commissioning phase of TopHAT with strongly variable
focus, and furthermore, conditions were non-photometric. Nevertheless,
we identified in the data a transit ingress event associated only with
the star \adsb.

Recent much higher quality observations using KeplerCam on the FLWO
\tsize{1.2} telescope have confirmed this, revealing the egress from the
transit on 2006 July 23, and later, on 2006 Aug 31. 
A small stamp from an FLWO \tsize{1.2} image that
clearly shows the two components of \adsab\ is exhibited in the right
panel of \figr{stamps}. The FLWO \tsize{1.2} data were
reduced using our own aperture photometry software (\fihat/\fiphot,
P\'al et al., in preparation). The photometry on individual frames was
then transformed to a selected reference frame using some one hundred
stars as comparison in the $23\arcmin\times23\arcmin$ field provided by
KeplerCam (we refer to this step as smooth magnitude transformation).
The resulting \lc\ for \adsb\ already showed a smooth transit curve
with very high S/N, but low amplitude systematic variations were
present. As a final correction, we
exploited the fact that \adsb\ has an ideal comparison star, namely
\adsa, with almost identical brightness, spectral type, and spatial
position. The data shown in the second and third panels of
\figr{hat_flwo48_lc} thus correspond to the differential magnitude 
of \adsb\ to \adsa\
after the smooth magnitude transformation procedure (shifted back to
an arbitrary $z \approx 9.53$ out-of-transit level). 

Unfortunately both FLWO \tsize{1.2} transit events were partial, thus we
seeked for further opportunities to observe a full transit.  Such an event
was observed recently in $I$-band, on 2006 Sep 5, using the 60/90/180cm
Schmidt telescope at the Piszk\'estet\H{o}\ mountain station of Konkoly
Observatory. Data are shown on the
bottom panel of \figr{hat_flwo48_lc}, and were reduced in a similar manner
to the FLWO \tsize{1.2} data.

The HATNet, TopHAT, FLWO \tsize{1.2} and Konkoly Schmidt data together 
span 1058 days, or 237
periods. Combining all the transit timings in these data we find
improved ephemerides:
$T_C (HJD) = E_0 + N_{tr} \cdot P$, where 
$E_0 = 2453984.397 \pm 0.009$, and $P=4.46529\pm0.00009$ (days), and 
$N_{tr}$ refers to the number of transits since the recent Konkoly
Schmidt observation that is a full transit observation, 
hence it has the best timing measurement.

\subsection{High-precision Spectroscopic Observations}
\label{sec:hispec}

In parallel to obtaining photometry follow-up data with the FLWO \tsize{1.2}
telescope, we obtained precise Doppler measurements of both \adsa\ and
B with the High Dispersion Spectrograph (HDS) on the \tsize{8.2} Subaru
telescope \citep{noguchi02}
to confirm the planetary nature of the HATNet photometric signal.
 
The HDS spectral format spans $3500 - 6100 \rm \AA$ over a mosaic of
two CCDs, and the $0\farcs 8$ slit yielded a resolution ($\lambda /
\Delta \lambda$) of 55000.  With only four observations we determined
that the brighter component, \adsa, had low rms velocity scatter while
component B exhibited velocity variations that were well
represented by a sine curve with an amplitude of about 60\ms, a
period equal to the photometric period, and a time of conjunction
consistent with that predicted from the transit ephemeris.
To improve phase coverage, nine observations of \adsb\ 
were subsequently obtained at W.~M.~Keck telescope over eight
nights in 2006 July using HIRES \citep{vogt94}. The spectrometer slit at
Keck is $0\farcs 86$ yielding a similar spectral resolution of about
55000 with a spectral format from $\sim 3200 - 8800 \rm \AA$.
The Subaru and Keck data were used to characterize more fully the
stellar properties of the system (\secr{stelprop}), as well
as to obtain a definitive radial velocity orbit (\secr{rvorbit}).

\section{Stellar Properties of the \ads\ system}
\label{sec:stelprop}

\subsection{SME analysis of the individual stars}
\label{sec:sme}

We have carried out spectral synthesis modeling of the iodine-free Keck
template spectra for both \adsa\ and B using the SME software
\citep{valenti96}, plus the wavelength ranges and atomic line data
described by \citet{valenti05}.  Results for \teff, \logg, \vsini\ and
\feh\ for both stars are shown in the first four lines of \tabr{stelpar}.  
Given the derived values of \teff\, the stars appear to fit the MK
classification of G0, rather than the F8 classification reported in
the literature.

\notetoeditor{This is the intended place of \tabr{stelpar}}
\begin{deluxetable}{lrr}
\tabletypesize{\scriptsize}
\tablecaption{
Summary of stellar parameters for \adsab.
\label{tab:stelpar}}
\tablewidth{0pt}
\tablehead{
	\colhead{Parameter} &
	\colhead{\adsa} &
	\colhead{\adsb}
}
\startdata
	\multicolumn{3}{c}{From SME analysis (\secr{sme})}\\
	\teff (K)			&	$6047\pm56$		&	$5975\pm45$\\
	\logg 				&	$4.13\pm0.10$	&	$4.45\pm0.06$\\
	\vsini (\kms)		&	$7.1\pm0.3$	&	$	2.2\pm0.2$\\
	\feh (dex)			&	$+0.12\pm0.05$	&	$+0.13\pm0.02$\\\hline

	\multicolumn{3}{c}{Simultaneous evolutionary track
		fitting(\secr{evo})}\\
	Mass (\msun)		&	$1.16\pm0.11$	&	$1.12\pm0.09$\\
	Radius (\rsun)		&	$1.23\pm^{0.14}_{0.10}$	& $1.15\pm^{0.10}_{0.07}$\\
	\loglstar (\lsun)	&	$0.26\pm0.15$	&	$0.18\pm^{0.17}_{0.14}$\\
	$M_I$ (mag)			&	$3.4\pm0.3$		&	$3.7\pm0.3$\\
	Age (Gyr)			&	\multicolumn{2}{c}{3.6}\\
	Z					&	\multicolumn{2}{c}{0.025}\\\hline

	\multicolumn{3}{c}{Other parameters}\\
	I (mag)				&	$9.035\pm0.05$	& $9.563\pm0.05$\\
	$\Delta I$ (mag)	&	\multicolumn{2}{c}{$0.53\pm0.03$}\\
	Mean Dist.~(pc)		&	\multicolumn{2}{c}{$139\pm^{22}_{19}$}\\
	Proj.~Sep.(AU)		&	\multicolumn{2}{c}{$1550\pm^{250}_{210}$}\\

\enddata
\end{deluxetable}

\subsection{Simultaneous evolutionary track fitting}
\label{sec:evo}

The two stars are very likely to comprise a physical pair,
for the following reasons:
First, they have common proper motion \citep{halbwachs86}.
Second, their relative apparent magnitudes compared to the absolute
magnitude expected for main-sequence stars with \teff\ and \logg\ given
in \tabr{stelpar} indicate that they are at a common distance.  This
also implies that their true tangential velocities are very similar,
and also that their real separation is small (projected separation
being $\sim$1550 AU).
Third, based on the six DS velocities for each star acquired over a
time interval of 234 days, they have common radial velocity within the
errors: 
$V_{\mathrm{rad,A}} = -3.43\pm0.32\, \mathrm{(rms)} \pm 0.14 \mathrm{(sys)} \kms$,
$V_{\mathrm{rad,B}} = -2.94\pm0.56\, \mathrm{(rms)} \pm 0.25 \mathrm{(sys)} \kms$.
Finally, they have identical metallicity within uncertainties. 
The physical companionship implies that the two stars are coeval, and 
provides the opportunity to perform simultaneous evolutionary
track fitting.

Furthermore, because of their common distance, the difference in their
apparent magnitudes yields an accurate measurement of the ratio of
their luminosities. The $I$ magnitudes of \adsa\ and \adsb\
respectively are 9.03 and 9.56. These values were determined by
cross-correlating the $I$ magnitudes for the \citet{landolt92} standard
stars with the $J$, $H$ and $K$ magnitudes of the 2MASS Point Source
Catalogue \citep[][]{skrutskie06}, and by deriving a weighted linear
regression between $J$, $H$, $K$ and $I$. The linear relation based on
$\sim400$ stars is very well defined, and has an rms of 0.06mag. The
$I$ magnitude difference, $\Delta I = 0.53\pm0.03$, is more accurately
determined than the individual magnitudes, because the stars are nearly
identical in spectral type and brightness so that systematic errors are
significantly canceled.

\begin{figure}[h]
\ifpdf
	\plotone{f3.pdf}
\else
	\plotone{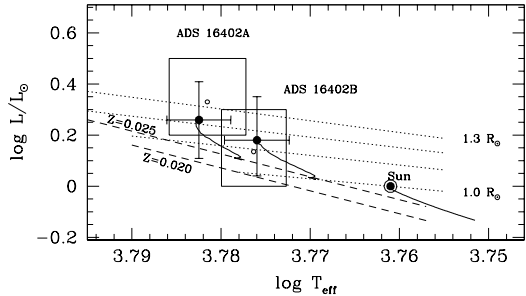}
\fi
\caption{
	Stellar evolution tracks (solid curved lines) for \adsa\ and \adsb\
	corresponding to the best mutual fit (filled circles with error
	bars).  The two stars are physically bound, are assumed to be
	coeval, and to have the same abundance, as determined by the SME
	analysis. Initial parameters from SME are shown with open circles
	with their respective (overlapping) error-boxes. The two tracks
	stop at 3.6 Gyr, so that the line joining their endpoints defines
	the local isochrone. The straight dashed lines plot the ZAMS for
	metallicity Z=0.020 and Z=0.025. For clarity the ZAMS for Z=0.030
	is not shown in the figure; it is displaced by the same amount
	above the ZAMS for Z=0.025 as the latter is above the ZAMS for
	Z=0.020. Dotted lines illustrate lines of equal radius. The
	position of the Sun and its track to its current age are also
	shown.
\label{fig:isochrones}}
\end{figure}

\notetoeditor{This is the intended place of \figr{isochrones}}

By exploiting the physical association of the two stars, their common
abundances, age, and their known luminosity ratio, we carried out a
simultaneous evolutionary track fitting. This in turn leads to a much
better determination of the parameters of \adsb\ than would be the case
if it were a single star.

Initial parameters (\figr{isochrones}, open circles) and their error
boxes (solid line rectangles in \figr{isochrones}) were based on the
atmospheric parameters obtained with the SME analysis of the Subaru
spectra. While the errors in \teff\ are taken directly from
\tabr{stelpar}, the absolute luminosities rely on the estimated gravity
and iteration with the mass determination, and this procedure is
imprecise. However, the ratio of the luminosities of the two stars is
well constrained, and is used in initializing the isochrone fitting to
both stars.

The value of \feh\ found from spectral synthesis corresponds to the
range of metal abundance by mass Z=0.020\---0.030. We have calculated
stellar evolution tracks of \adsa\ and \adsb\ using Z in this range
(\figr{isochrones}). The straight line that would connect the centers
of the two error boxes is steeper than the theoretical isochrones (the
straight line that would connect the end tips of the two evolutionary
tracks) for any Z in the range Z=0.020\---0.030. All possible solutions
are some sort of compromise within the observational error boxes,
imposed by the requirement for equal age, and with the obvious
additional constraint that each track must begin on the ZAMS
appropriate to the chosen value of Z.  For different Z's, different
mass ranges sweep the two error boxes \--- smaller ones for Z=0.020 and
progressively larger for higher Z, with a largely normal distribution. 
The observational error box of \adsb\ from the SME analysis
(\figr{isochrones}) overlaps with the ZAMS for any Z, and naturally,
the star cannot lie below the ZAMS for the accepted abundance, so this
constrains the possible solution from the lower bounds of
\logg\ and \logl. The distribution of possible stellar radii is
asymmetric due to these constraints and this leads later to asymmetric error
bars of the planetary radius.

Our best fit theoretical isochrone is for Z=0.025, and roughly bisects
the line connecting the centers of the SME error boxes. This gives
masses of
$1.12\pm 0.09$ \msun\ and
$1.16\pm0.11$ \msun\, and radii of 
$1.15\pm^{0.10}_{0.07}$\rsun\ and 
$1.23\pm^{0.14}_{0.10}$\rsun, 
for \adsb\ and \adsa\ respectively, and an age of 3.6 Gyrs for the system
(see Table 1), and we use these in the further analysis.

The absolute $I$ magnitude of \adsb\ is $M_I = 3.74 \pm 0.3$ mag (with
bolometric correction of $+0.55$ mag).  From its apparent $I$ magnitude
of $9.56\pm0.06$mag (see above), and by neglecting reddening
(being a near-by source at $b=20\arcdeg$) we derive a distance of
$146\pm^{24}_{21}$pc.  The same calculation for \adsa\ yields
$131\pm^{21}_{17}$pc. We estimate the distance of the \adsab\ system as
the average of the two: $139\pm20$pc. The 11\farcs2 separation of the
two components then implies a projected separation of $1550 \pm 250$
AU.

All derived stellar parameters, along with absolute and apparent $I$
magnitudes, magnitude difference, mean distance, and projected
separation of the stellar components of \ads\ are given in
\tabr{stelpar}.

\section{Spectroscopic Orbit of \adsb}
\label{sec:rvorbit}

\notetoeditor{This is the intended place of \tabr{radvel}}
\begin{deluxetable}{lrrr}
\tabletypesize{\scriptsize}
\tablewidth{0pt}
\tablecaption{
	\label{tab:radvel}
	Radial Velocities for \adsb.
}
\tablewidth{0pt}
\tablehead{
	\colhead{HJD - 2453000} &
	\colhead{RV} &
	\colhead{Uncert.} &
	\colhead{Observatory} \\
	\colhead{(days)} &
	\colhead{(\ms)} &
	\colhead{(\ms)} &
	\\
}
\startdata
    897.11285 &  -2.89  &   5.00  & Subaru \\
    899.12678 &  52.53  &   4.77  & Subaru \\
    900.12008 & -34.51  &   4.65  & Subaru \\
    901.10260 & -37.51  &   4.88  & Subaru \\
    927.06848 & -37.76  &   3.84  & Keck   \\
    927.96558 & -28.27  &   3.90  & Keck   \\
    931.03690 & -12.10  &   3.83  & Keck   \\
    931.94061 & -46.59  &   3.92  & Keck   \\
    932.03590 & -42.95  &   3.83  & Keck   \\
    932.99985 &   3.53  &   3.85  & Keck   \\
    933.92455 &  66.16  &   4.04  & Keck   \\
    934.90368 &  49.20  &   4.92  & Keck   \\
    934.90692 &  49.21  &   4.00  & Keck   \\
\enddata                         
\end{deluxetable}

The Subaru HDS Doppler observations were obtained through an Iodine
absorption cell (in front of the spectrometer slit for this observing
run) to provide a fiducial wavelength scale and to model the
instrumental point spread function \citep{kambe02}. 
%
%
We measured the star's radial velocity from each high-resolution
spectrum, using a synthetic stellar template that was created by
modifying a close-match spectrum from a library of high resolution
template spectra obtained at Keck \citep{johnson06}. Doppler velocities
at Keck are also modeled from observations taken through an Iodine
cell, however, a standard template observation (obtained without the
Iodine cell) was used for the reference spectrum \citep{butler96}.
Exposure times at both Subaru and Keck were typically 500 s,
yielding an S/N of about 150.

Emission in the core of the Ca H\&K lines was measured in each of the
Keck spectra to assess the chromospheric activity of \adsa\ and B. With
five spectra for \adsa, we measure an average value for $S_{HK} = 0.16$
corresponding to $\log R'_{HK} = -4.923$ for the A component and an
average value of nine spectra yields $S_{HK} = 0.14$ and $\log R'_{HK}
= -5.03$ for the B component.  These values indicate low chromospheric
activity. Our estimate for intrinsic astrophysical velocity jitter is
$3.7\ms$ for both stars \citep{wright05}.

\begin{figure}[h]
\ifpdf
	\plotone{f4.pdf}
\else
	\plotone{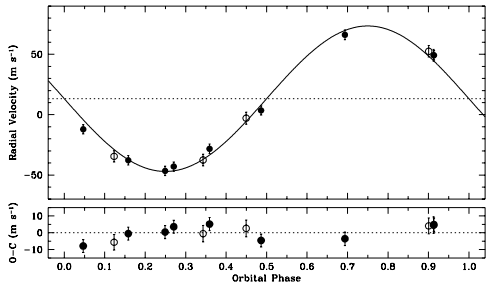}
\fi

\caption{
	Radial Velocities from Subaru and Keck, folded with the period of
	4.46529 days. Phase 0 corresponds to the predicted time of
	mid-transit from the orbital ephemeris given in text. Open circles
	represent Subaru data (after the $14.5\ms$ shift described in the
	text, whereas filled circles show Keck measurements.
\label{fig:PRVdata}}
\end{figure}

Radial velocity measurements from Subaru and Keck were used together to
derive an orbital fit. The reference velocities were those from Keck, and
an offset ($\Delta_v$) between Subaru and Keck was also fitted for in
the orbital solution.  Velocity measurements are given in \tabr{radvel}
along with observation times and errors. The velocity offset of
$\Delta_v = 14.5\ms$ has been already applied to the Subaru data shown in
the table, and the expected velocity jitter of $3.7\ms$ has already
been added in quadrature to the uncertainties listed in the table. The
photometric period of $4.46529$ days, and the mid-transit time from the
photometric ephemerides were fixed in this model. We also assumed zero
eccentricity based on the theoretical expectation of circular orbits
for periods as short as this. In summary, the fitted parameters were
the $\gamma$ velocity of the Keck data points, the $K$ semi-amplitude,
and the $\Delta_v$ velocity difference between Keck and Subaru. To
determine the uncertainties, 100 Monte Carlo fitting trials were run.
In each trial, the best fit Keplerian model was subtracted from the
velocities and the residuals were shuffled and added back with
replacement to the theoretical velocities. From these trials, we find a
velocity semi-amplitude, $K = 60.3 \pm 2.1\ms$, an rms of $5.11\ms$ and
a reduced $\chi^2 = 1.54$.  The final fit is over-plotted on the RV
data in the upper panel of \figr{PRVdata}. The lower panel shows the
residuals.

Adopting a stellar mass of
$1.12 \pm 0.09 \msun$, we derive 
$a_{\rm rel} = 0.0551\pm0.0015$ AU and
$\msini = 0.53 \pm 0.04\mjup$.
Using an inclination 
$\ipl = 85\fdg9 \pm 0\fdg8$
(\secr{PlanetProperties} below), this yields a 
planetary mass $\mpl = 0.53 \pm 0.04\mjup$.
The orbital elements and uncertainties are summarized in \tabr{orbit}.

There is a small but non-negligible systematic trend in the residuals
of the orbital fit (lower panel of \figr{PRVdata}). We note that the
fit is improved if the eccentricity and $\omega$ are allowed to float,
and we get 
$e = 0.09\pm0.02$, and
$\omega = 80\fdg7 \pm 7\fdg6$ with
rms of $3.2\ms$ and reduced $\chi^2 = 0.53$. The non-zero eccentricity
is only suggestive, however, because of the i) small number of
data-points, and ii) the unrealistically small reduced $\chi^2$.
Nevertheless, implications of this finding are potentially very
important, and we elaborate further on them in the discussion
(\secr{discus}).

\notetoeditor{This is the intended place of \figr{PRVdata}}

\notetoeditor{This is the intended place of \tabr{orbit}}
\begin{deluxetable}{lr}
\tablecaption{
	\label{tab:orbit}
	Parameters of the \hatp\ planetary system.
}
\tablewidth{0pt}
\tablehead{
	\colhead{Parameter} &
	\colhead{Value}\\
} 
\startdata
Period (d)					&	$4.46529\pm0.00009$\tablenotemark{a}\\
${\rm T}_{\rm mid}$ (HJD)	&	$2453984.397\pm0.009$\tablenotemark{a}\\
$\omega$ (deg)				&	0\tablenotemark{a}	\\
ecc							&	0\tablenotemark{a}	\\
K$_1$ (\ms)					&	$60.3\pm2.1$		\\
$a_{\rm rel}$ (AU)			&	$0.0551\pm0.0015$	\\
\msini (\mjup)				&	$0.53\pm0.04$		\\
\ipl (deg)					&	$85.9\pm0.8$		\\
\mpl (\mjup)				&   $0.53\pm0.04$		\\
\rpl (\rjup)				&	$1.36\pm^{0.11}_{0.09}$
\enddata
\tablenotetext{a}{Fixed in the orbital fit.}
\end{deluxetable}

\section{Excluding blend scenarios}
\label{sec:blend}

\begin{figure}[h]
\ifpdf
	\plotone{f5.pdf}
\else
	\plotone{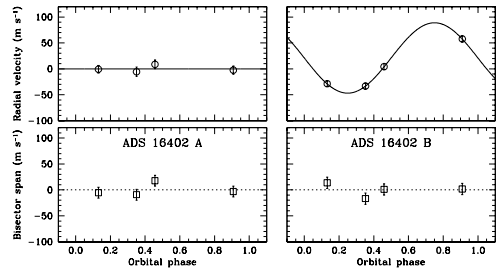}
\fi

\caption{
	Comparison between the relative radial velocities and bisector
	spans of \adsa\ (left panel) and \adsb\ (right panel), based on the
	Subaru spectra. \adsa\ shows no variation in either the velocities
	or the bisectors spans (left panels), while \adsb\ shows obvious
	velocity variations but does not show the variation in the bisector
	spans that would be expected if the velocity signal were the result
	of a blend with an eclipsing binary.
\label{fig:bisec}}
\end{figure}

We investigated the possibility that the measured radial-velocity
variations are due not to a planetary companion but rather to
distortions of spectral line profiles arising from contamination of the
spectrum by the presence of an unresolved binary companion to \adsb,
with an orbital period of $\sim$4.4653 days and an amplitude of several
tens of \kms. However, this would give rise to time-varying asymmetry
in the spectral line bisectors \citep[see, e.g.,][]{santos02,
torres05}.  We have searched for such line bisector variations in our
Subaru spectra and find no evidence for them.
Comparison between the relative radial velocities and bisector spans of
\adsa\ and B is shown in \figr{bisec}. The results are based on the
Subaru spectra only, because they were obtained on the same nights for
both stars. Line bisectors for each spectrum were computed from the
cross-correlation function averaged over all orders, which is
representative of the average spectral line profile of the star. The
cross-correlations were performed against a synthetic spectrum matching
the properties of the stars. The bisector spans were then computed as
the velocity difference between points chosen near the top and bottom
of the bisectors \citep[see][]{torres05}. \adsa\ shows no variation in
either the velocities or the bisectors spans. \adsb, on the other
hand, shows obvious velocity variations but does not show the
concomitant change in the bisector spans that would be expected if it
were the result of a blend with an eclipsing binary. 
If the light and velocity variations were the result of a blend with
an eclipsing binary, the bisector spans would be expected to vary with
a similar amplitude as the velocities \citep[see, e.g.,][]{queloz01}.
That scenario is therefore ruled out by these observations.

\notetoeditor{This is the intended place of \figr{bisec}}

We have also modeled the light curve as a blend following
\cite{torres04b}, subject to all available observational constraints.
Specifically, we calculated model light curves for a three-star system
consisting of a slightly evolved early G star primary plus a later-type
secondary and a tertiary which together comprise an eclipsing binary,
with the requirement that the brightness of the secondary be no more 
than $\sim$10\% that of the primary (to have escaped detection in the
DS spectra) and that the light curve of the eclipsing binary,
attenuated by the much brighter primary, should be indistinguishable
from the observed light curve in the FLWO \tsize{1.2} data.
However, we have found it impossible to construct an acceptable light
curve in this fashion.

Based on the above, we conclude that \adsb\ is orbited by a transiting
object that gives rise to the observed photometric and radial velocity
variations. From the 60\ms\ amplitude of the radial velocity orbit
(\secr{rvorbit}), we conclude that the object must have mass about 0.53
\mjup\, and hence must be an extrasolar planet.

\section{The Transiting Planet \hatp\ and its Properties}
\label{sec:PlanetProperties}

In order to determine more precisely the physical characteristics of
\hatp\ and its orbital parameters, we have carried out model fitting
of the different sets of photometric data shown in the four panels of
\figr{hat_flwo48_lc}. A simple goodness-of-fit was determined against
the predicted flux values evaluated with the algorithm of
\citet{mandel02}. We used the values for the stellar radius and mass
(and their uncertainties) determined above, and coefficients for
stellar limb darkening for the Sloan $z$-filter by Claret (2004).
First we fitted the original discovery data from HATNet with the
appropriate correction for the flux of the binary star (unresolved
on the HATNet images): the \rpl\ = $1.42$\rjup\ and $\ipl = 85\fdg4$,
have large uncertainties, as anticipated, and might be
subject to large systematic errors. We then fitted to the recent
Konkoly Schmidt observations of a complete \hatp\ transit in $I$-band: the
values are \rpl\ = $1.31 \pm 0.07$\rjup\ and $\ipl = 86\fdg0 \pm 1\fdg0$,
and a precise estimate of transit duration of $d = 0.11671$ days
and center of transit $T_c = 3984.3967$. The latter value
allows us to estimate accurately the $T_c$ of the transit we observed
just an orbital period earlier (Aug 31) at FLWO, where the beginning of
the ingress is missing (see \figr{hat_flwo48_lc}, third panel from the
top). Combining both $z$-band transit curves from FLWO with a known
$T_c$ allows an independent fit: we get \rpl\ = $1.36 \pm 0.05$\rjup\ 
and $\ipl = 85\fdg9 \pm 0\fdg8$. The fit is better because the
quality and amount of photometric data is higher, and the limb
darkening in $z$-band is less. The fits to all data sets give us the same
planet parameters within the error; henceforth we use our best fit with
the FLWO $z$-band transits.

The result for the planetary radius is \rpl\ =
$1.36\pm^{0.11}_{0.09}$\rjup, where the above statistical errors from
the fit and systematic errors from the uncertainties of the stellar
radius and mass ($\pm^{0.09}_{0.07}$), have been added in quadrature. 
Due to the asymmetric error bars on the stellar radius (see above) we
have asymmetric error bars on \rpl\ too. The orbital inclination is
$\ipl = 85\fdg9 \pm 0\fdg8$. The radius, orbital inclination, and
derived mass of the planet are included in \tabr{orbit}.

An even better \lc\, with e.g.~the {\em Hubble Space Telescope}, 
and a more detailed analysis of
the parameters of the stellar pair could reduce that error
considerably. The results of this on-going work will be reported in a
forthcoming paper, as soon as the multi-band observations enable us to
refine the binary system parameters.

\section{Discussion}
\label{sec:discus}

\begin{figure}[h]
\ifpdf
	\plotone{f6.pdf}
\else
	\plotone{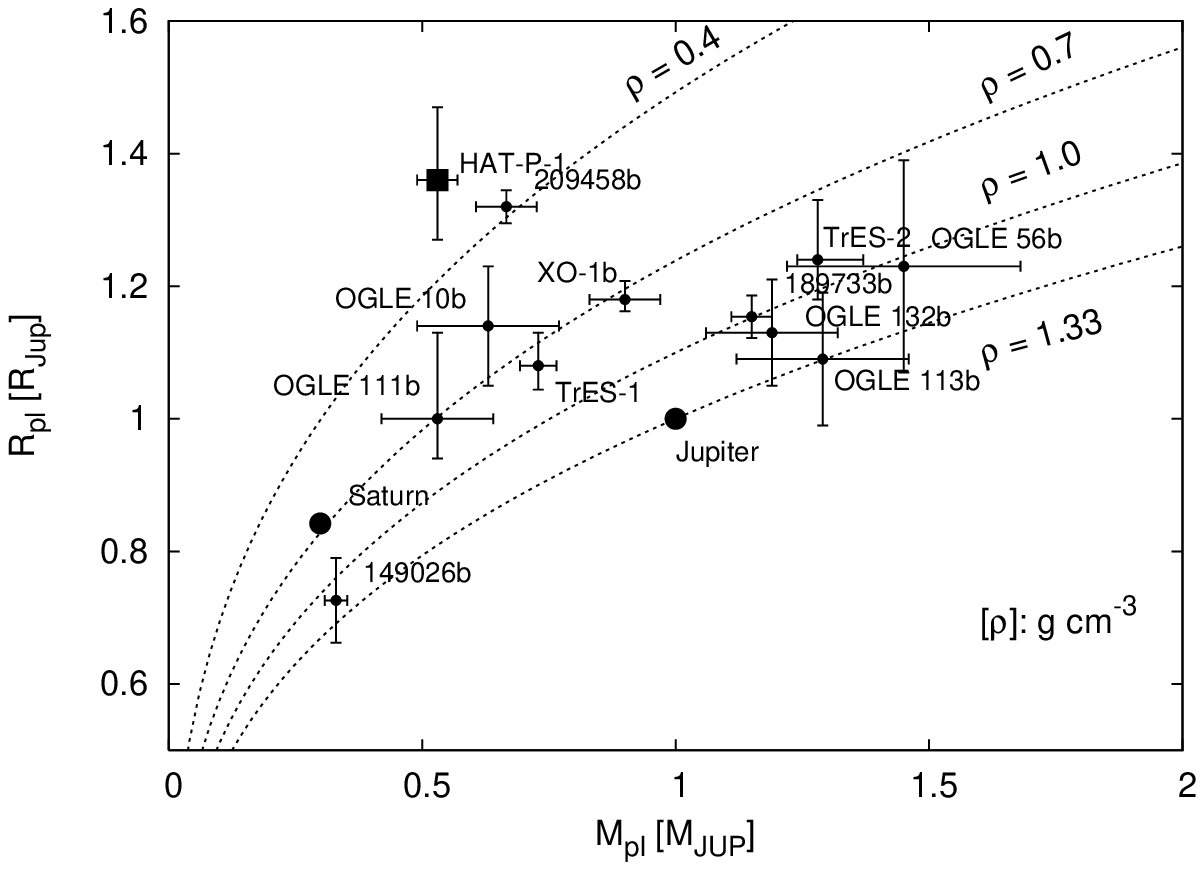}
\fi

\caption{
	Mass-radius diagram depicting known transiting exoplanets, plus
	Saturn and Jupiter (for comparison). Sources for the mass and
	radius values (in this order) are listed after the name of the
	planet;
	\mbox{\hatp}: this work,
	\mbox{HD 209458b}: \citet{laughlin05}, \citet{knutson06},
	\mbox{HD 189733b}: \citet{bakos06},
	\mbox{OGLE-TR-111b}: \citet{pont04}, 
	\mbox{OGLE-TR-10b}: \citet{santos06},
	\mbox{OGLE-TR-132b}:  \citet{moutou04},
	\mbox{OGLE-TR-56b}: \citet{torres04a},
	\mbox{OGLE-TR-113b}: \citet{bouchy04,konacki04},
	\mbox{TrES-1}: \citet{sozetti04},
	\mbox{HD 149026}: \citet{sato05}, \citet{dc06a},
	\mbox{XO-1b}: \citet{holman06},
	\mbox{TrES-2}: \citet{odonovan06}.
\label{fig:exoplanet_m_r}}
\end{figure}

\figr{exoplanet_m_r} shows the location of \hatp\ in a plot of
planetary radius versus planetary mass for the now 11 known transiting
exoplanets, plus Jupiter and Saturn.  It is immediately seen that
\hatp\ is in an extreme position. It has a radius apparently as large
as or larger than the largest known exoplanet, \mbox{HD 209458b}, and at the
same time it has a mass that is significantly smaller.  Thus, it has a
lower mean density than any other known planet. Until now, \mbox{HD 209458b}
was the only significantly anomalous transiting planet. Thus,
\citet{guillot05} noted that all of the transiting planets known at
that time, with the exception of \mbox{HD 209458b}, fall within a mass-radius
relation derived from modeling the interiors of strongly irradiated
planets. All of the transiting planets detected subsequently, except
\hatp, also fit the same relation. Among them, OGLE-TR-10b had been
thought to be similarly anomalous \citep{konacki05}, but recent
evidence \citep[][M.~Holman, private communication]{holman05,santos06}
points to a smaller planetary radius and normal density. Similarly, the
XO-1b discovery paper \citep{mccullough06} derives a larger radius than the
value later refined by \citep{holman06}.

\notetoeditor{This is the intended place of \figr{exoplanet_m_r}}

\mbox{HD 209458b} has a radius some 10 to 20 percent larger than expected,
given its age, mass, and expected temperature including effects of
irradiation. Possible solutions include: (a) a fraction of a percent of
the stellar flux irradiating \mbox{HD 209458b} could be transported to deep
layers of order 100 bar, and thereby keep the planet's radius puffed up
\citep{showman02}; (b) there is a small eccentricity of its orbit,
forced by the presence of a so-far undetected planetary companion and
leading to internal tidal heating \citep{bodenheimer03}, and (c) HD
209458b might be in a Cassini State---a dynamical state where its
rotation axis lies close to the plane of its orbit \citep{winn05a},
leading to large obliquity tides that cause heating at significant
depth leading to the planet's large radius. A natural question about
the first of these mechanisms is why all close-in ``hot Jupiters'' are
not similarly bloated by that mechanism. A problem with the last two
mechanisms is that both are {\em ad hoc}; they require unusual
circumstances for which there is no independent evidence. 

With the discovery of another planet, \hatp, which has a radius
similar to and probably even larger than \mbox{HD 209458b}, and an even
smaller density, mechanisms requiring unusual circumstances seem less
likely.  Moreover, for \mbox{HD 209458b}, explanation (b) appears ruled out
because Doppler data \citep{laughlin05b,winn05b} 
and transit timing variations
data \citep{miller05} have excluded companion planets capable of
enacting it.

In the case of \hatp, it remains possible that the system has a second
planet capable of forcing its eccentricity and leading to internal
tidal heating. Our current orbital solution (\secr{rvorbit}) suggests
$e=0.09 \pm 0.02$. If confirmed, such eccentricity would be a strong
indication for the presence of a third body (second planet), because
otherwise tidal dissipation would have circularized \hatp's orbit
in $\sim 225 (\frac {Q}{10^6})$~Myr. The dissipation factor, $Q$, is
poorly known for ``hot Jupiters'', but is expected to lie between the
values for Jupiter and solar-like stars ($10^5$ to $10^6$). With
forced orbital eccentricity, $e=0.09$, \hatp\ will be subject to
internal tidal heating due to an energy dissipation rate
\begin{displaymath}
\frac{dE}{dt} = \
\frac{63}{4}\frac{e^{2}n}{Q}\left(\frac{R_{p}}{a}\right)^{5}\frac{G\mstar^{2}}{a}
\end{displaymath}
where $G$ is the gravitational constant, $M_{\star}$ is the mass
of the star, $a$ is the semimajor axis, $n$ is the mean motion,
and $Q$ is the specific dissipation factor.
With $e = 0.09$, the rate of energy dissipation in \hatp\ would be
$3.0\times 10^{26} erg~s^{-1}$, corresponding to energy dissipation per
unit Jupiter mass of $5.7 \times 10^{26} erg~ s^{-1}\mjup^{-1}$.
This rate is almost identical to the dissipation rate per unit Jupiter
mass of $5.8\times 10^{26} erg~s^{-1} M_{J}^{-1}$ required to achieve a
20\% increase in \rpl\ for \mbox{HD 209458b} (no core model) and explain its
anomalous low density \citep{bodenheimer03}. Therefore we conclude that
if the orbital eccentricity is confirmed, the anomalous \rpl\ and low
density of \hatp\ could be easily explained by tidal heating. The
perturbing second planet would be detectable by both transit timing
variations and Doppler velocities.

For now, however, all that can be said is that bloated planets are not
that unusual, and, since a second perturbing planet is lacking in HD
209458b's case, perhaps some other explanation should be sought that
can explain such planets more generally.

\acknowledgments
Operation of the HATNet project is funded in part by NASA grant
NNG04GN74G.  
Support for program number HST-HF-01170.01-A to G.~\'A.~B.~was provided by
NASA through a Hubble Fellowship grant from the Space Telescope Science
Institute, which is operated by the Association of Universities for Research
in Astronomy, Incorporated, under NASA contract NAS5-26555.
G.~K.~wishes to thank support from Hungarian Scientific Research
Foundation (OTKA) grant K-60750.
We acknowledge partial support from the Kepler Mission under NASA
Cooperative Agreement NCC2-1390 (D.~W.~L., PI).
G.~T.~acknowledges partial support from NASA Origins grant NNG04LG89G.
We thank Akito Tajitsu for his expertise and support of the Subaru HDS
observations. We thank graduate students M.~Ohmiya, S.~Robinson and
K.~Peek for help collecting data at Subaru and Keck.
The Keck Observatory was made possible by the generous financial
support of the W.~M.~Keck Foundation. D.~A.~F is a Cottrell Science
Scholar of Research Corporation. We acknowledge support from NASA grant
NNG05G164G to DAF.
We would like to thank Carl Akerlof and the de-commissioned ROTSE-I
project the generous loan of some of the lenses and CCDs that we use
for operating HATNet.

We owe special thanks to Emilio Falco, Dan Fabricant, James Moran and
Antony Schinckel for their help in establishing and operating the
HATNet stations at FLWO and SMA.
G.~\'A.~B.~wishes to thank the support
given by telescope operators Mike Calkins and Perry Berlind in the
operation of the FLWO HATNet station. We also thank Gergely G\'alfi for
useful discussions.

This publication made use of the VizieR interactive catalogue
\citep{ochsenbein00} at CDS, Strasbourg, and the 2MASS catalogue.




\end{document}